\documentclass{elsart}

\usepackage{natbib}

\usepackage{amssymb}

\usepackage{graphicx} 
\usepackage[figuresright]{rotating}

\begin{document}

\begin{frontmatter}


\title{Migration of Small Bodies and Dust to Near-Earth Space} 
\author{S.I. Ipatov} 
\ead{siipatov@hotmail.com}
\corauth[cor1]{}
\address{Catholic University of America, Washington DC, USA; Institute of Applied Mathematics, 
Moscow, Russia}

\author{J.C. Mather} 
\ead{John.C.Mather@nasa.gov}
\corauth[cor1]{}
\address{NASA/Goddard Space Flight Center, Greenbelt, MD, USA}








\begin{abstract}

The orbital evolution of Jupiter-family comets (JFCs), 
resonant asteroids, and asteroidal, trans-Neptunian, and cometary dust particles 
under the gravitational influence of planets
was integrated.  For dust particles we also 
considered radiation pressure, Poynting-Robertson drag, and solar wind drag. 
The probability of a collision of one  former JFC with a 
terrestrial planet can be greater than analogous total probability for thousands 
other JFCs.
If those former JFCs that got near-Earth object (NEO) orbits for millions of years
didn't disintegrate during this time, there could be many extinct comets among NEOs.
The maximum probability of a collision of an asteroidal or cometary dust particle with the Earth during its lifetime was for diameter $d$$\sim$100 microns. At $d$$<$10 micron, the collision probability of a trans-Neptunian particle with the Earth during a lifetime
of the particle was less than that for an asteroidal particle  by only a factor of several.

\end{abstract}


\end{frontmatter}

\section{INTRODUCTION}

Duncan et al. (1995) and Kuchner et al. (2002)  investigated the migration of TNOs
to Neptune's orbit, and  Levison and Duncan (1997)  studied the 
migration from Neptune's orbit to Jupiter's orbit. 
Ipatov (2002), Ipatov and Mather (2003, 2004a-b) integrated the orbital evolution
of Jupiter-family comets (JFCs) under the gravitational influence of planets,
paying the main attention to their migration to the near-Earth space and
collisional probabilities with the terrestrial planets. 
In section 2 of the present paper we summarize our investigations
of this problem. 

Liou et al. (1996), Liou and Zook (1999), Gorkavyi et al. (2000), 
Ozernoy (2001), Moro-Martin and Malhotra (2002, 2003) considered the
migration of dust particles from the trans-Neptunian belt, Liou et al. (1999) from 
Halley-type comets, Liou et al. (1995) from Comet Encke, and Reach et al. (1997), 
Kortenkamp and Dermott (1998), and Grogan et al. (2001) from asteroid families and 
short-period comets. Further references are presented in the above papers and by Dermott et al. (2002). 

Ozernoy (2001) considered 1 $\mu$m and 5 $\mu$m 
particles while constructing the brightness of a disk of asteroidal, cometary, and trans-Neptunian
(kuiperoidal) dust particles which fit the COBE/DIRBE data and concluded that 
the trans-Neptunian dust contributes as much as 1/3 of the total 
number density near the Earth. Liou et al. (1995) showed that the observed shape 
of the zodiacal cloud can be accounted for by a combination of about 1/4 to 1/3 
asteroidal dust and about 3/4 to 2/3 cometary dust.
The mass distribution of dust particles falling to the Earth 
peaks at about 200 $\mu$m in diameter (Gr\"un et al., 1985).

In the present paper we consider a wider range of masses 
(including particles up to 1000 $\mu$m) of 
asteroidal and cometary dust particles than the aforementioned authors
and Ipatov et al. (2004a). 
We pay more attention to the migration of trans-Neptunian dust particles
to the near-Earth space
than Liou and Zook (1999) and Moro-Martin and Malhotra (2002, 2003).

\section{ORBITAL EVOLUTION OF JUPITER-FAMILY COMETS} 

Ipatov (2002), and Ipatov and Mather (2003, 2004a-b) 
investigated migration of JFCs to the near-Earth space.
 Note that the paper by Ipatov and Mather (2004a) was based on more recent runs than other 
above papers. References on papers by other scientists can be found in our previous papers.
We integrated the orbital evolution of $\sim$30,000 JFCs and 1300 resonant main-belt asteroids
under the gravitational influence of planets during dynamical lifetimes of the 
small objects.  We omitted the influence of Mercury (except for Comet 2P/Encke) and Pluto and used the Bulirsch-Stoer and symplectic methods (BULSTO and RMVS3 codes) from the integration package of Levison and Duncan (1994). 
In two series of runs, initial orbits were close to those of several (10 and 20) real JFCs, and
in each of other series they were close to the orbit of a single real JFC  
(2P, 9P, 10P, 22P, 28P, 39P, or 44P). In our runs, planets were considered as material points, so 
literal collisions did not occur.  However, using the orbital elements sampled with a 500 yr 
step, we calculated the mean probability  $P$ of collisions (see our previous papers for details). 

In our runs, some former JFCs moved in typical near-Earth object (NEO) orbits, and a few of them got orbits with aphelion distance $Q$$<$3 AU for millions of years. The main portion of the probability of collisions of former JFCs with the 
terrestrial planets was due to a small ($\sim$0.1 \%) portion of objects that moved during several Myrs in orbits with $Q$$<$4.2 AU. 
The mean collision probabilities of JFCs  with the terrestrial planets can differ 
for different comets by more than two orders of magnitude. 
The ratio of the mean probability of a JFC with semi-major axis $a$$>$1 AU with a planet to the mass of the planet was greater for Mars by a factor of several than that for Earth and Venus. 
Four considered former JFCs even got inner-Earth orbits (with $Q$$<$0.983 AU) or Aten orbits ($a$$<$1 AU, $Q$$>$0.983 AU) for Myrs. One former JFC got Aten orbits  during $>$3 Myr
and inner-Earth orbits  for $\sim$10 Myr, but  probabilities of its collisions with  Earth and Venus were greater than those for $10^4$ other former JFCs.  The number
of JFCs considered was greater  than that 
considered by Bottke et al. (2002) by an order of magnitude, 
so that is why these scientists didn't obtain
orbits with $a$$<$2 AU. Note that Ipatov (1995) obtained migration of JFCs into inner-Earth
and Aten orbits using the method of spheres.

Former JFCs can get typical asteroidal orbits, but even less often than NEO orbits.
After 40 Myr one considered object (with initial orbit close to that of Comet 88P) 
got $Q$$<$3.5 AU, and it moved in orbits with $a$=2.60-2.61 AU, perihelion distance 1.7$<$$q$$<$2.2 AU, 3.1$<$$Q$$<$3.5 AU, eccentricity $e$=0.2-0.3, and inclination 
$i$=5-10$^\circ$ for 650 Myr. Another object (with initial orbit close to that of Comet 94P) moved in orbits with $a$=1.95-2.1 AU, $q$$>$1.4 AU, $Q$$<$2.6 AU, 
$e$=0.2-0.3, and $i$=9-33$^\circ$ for 8 Myr (and it had $Q$$<$3 AU for 100 Myr). In our opinion, it can be possible that Comet 133P (Elst-Pizarro) moving in a typical asteroidal orbit was earlier a JFC and it circulated its orbit also due to non-gravitational forces. 

The results obtained by the Bulirsh-Stoer method (BULSTO code) with the 
integration step error less than $\varepsilon$, where 
$10^{-9}$$\le$$\varepsilon$$\le$$10^{-8}$ 
and $\varepsilon$$\le$$10^{-12}$, 
and by a symplectic method (RMVS3 code) at integration step $d_s$$\le$10 days were usually similar.
The difference at these three series was about the difference at small variation of $\varepsilon$ or $d_s$. In the case of close encounters with the Sun (i.e., for Comet 2P, Comet 96P, and the 3:1 resonance with Jupiter), the values of collision probability 
$P_S$ with the Sun obtained by BULSTO and RMVS3 and at different $\varepsilon$ or $d_s$ were different, but all other results were usually similar,
as most bodies didn't move long after close encounters. 

In comparison with Ipatov and Mather (2003, 2004a-b), we made 
aditionally a series of runs for Comet 2P at $d_s$$\le$3 days. Earlier runs were made for $d_s$=10 and $d_s$=30 days.
The obtained values of $P_S$ were similar for different $d_s$; they were 0.98, 0.99, 0.99, 0.96, and 0.99 for $d_s$ equal to 0.1, 0.3, 1, 3, and 10 days, respectively. Note that for BULSTO $P_S$=0.88 at $\varepsilon$=$10^{-13}$ 
and $\varepsilon$=$10^{-12}$ ($P_S$ was even smaller for larger $\varepsilon$),
i.e., $P_S$ was smaller than for RMVS3. In all runs the minimum values of times elapsed up to
a collision of an object with the Sun were about 40-60 Kyr, but the maximum values of times 
varied considerably in different runs (from 1 Myr to 400 Myr).
At $d_s$=3 days, a lifetime of one object was 400 Myr, and it moved
on Inner-Earth, Aten, Apollo, and Amor orbits during 2.5, 2.2, 44.9, and 80.8 Myr,
respectively. At $t$=6.5 Myr this object got an orbit with $e$=0.03 and $a$=1.3 AU,
and then until 370 Myr  the eccentricity was less than 0.4 and often was even less than 0.2.
The probability of a collision of this object with the Earth was about 1, and it was
greater than that for all other 99 objects in that run
by two orders of magnitude.

Ipatov and Mather (2004a) showed that during the accumulation of the giant planets the total mass of icy bodies delivered to the Earth could be about the mass of water in Earth's oceans. Many Earth-crossing objects can move in highly eccentric ($e$$>$0.6) orbits and, probably, most of 1-km objects in such orbits have not yet been discovered. 
If one observes former JFCs in NEO orbits, then most of them could have already moved in such orbits for millions (or at least hundreds of thousands) of years. Some former comets that have moved in typical NEO orbits for millions or even hundreds of millions of years, and might have had multiple close encounters with the Sun, could have lost their typically dark surface material, thus brightening their low albedo and assuming the aspect typical of an asteroid (for most observed NEOs, the albedo is greater than that for comets). 
On the contrary, Napier et al. (2004) suggested that the surfaces of innert comets became
extremely dark and most of such comets became invisible. 
If many of extinct comets disintegrated, then there can be many mini-comets in the near-Earth space, and the Tunguska comet could be one of them.
At least one of the below conclusions follow from our runs: 1) the portion of 1-km former trans-Neptunian objects (TNOs) among NEOs can exceed several tens of percent, 2) the number of TNOs migrating inside solar system could be smaller by a factor of several than it was earlier considered, 3) most of 1-km former TNOs that had got NEO orbits disintegrated into mini-comets and dust during a smaller part of their dynamical lifetimes if these lifetimes are not small. 

\section{Models for migration of dust particles}

    Using the Bulirsh--Stoer method of integration, we investigated
the migration of dust particles under the influence of planetary gravity
(excluding Pluto for asteroidal and cometary particles), radiation pressure, Poynting--Robertson drag, and solar wind drag.
The initial positions and velocities of the asteroidal particles 
were the same as those of the first $N$ numbered main-belt asteroids (JDT 2452500.5),
i.e., dust particles are assumed to leave the asteroids with zero
relative velocity. 
For the runs marked with * in Table 1, we considered next $N$ asteroids. 
The initial positions and velocities of the 
trans-Neptunian particles 
were the same as those of the first TNOs (JDT 2452600.5),
and our initial data were different from those in previous papers.
In each run we took $N$$\le$250 particles, because for $N$$\ge$500 
the computer time per calculation for one particle was several times greater 
than for $N$=250. 
In the main series of runs, the initial positions and velocities of the cometary particles were the same as 
those of Comet 2P Encke ($a$$\approx$2.2
AU, $e$$\approx$0.85, $i$$\approx$12$^\circ$). 
We considered Encke particles starting near perihelion 
(runs denoted as $\Delta t_o$=0), 
near aphelion ($\Delta t_o$=0.5), and when the comet had orbited for $P_a/4$ 
after perihelion passage, where $P_a$ is the period of the comet
(such runs are denoted as $\Delta t_o$=0.25).
Variations in time $\tau$ when perihelion was passed was varied 
with a step 
$0.1$ day for series with $N$=101 and with a step 1 day for series with $N$=150,
so for $N$=101 initial positions of Encke particles were more compact.
We also studied migration of dust particles started from Comet 10P/Tempel 2 ($a$$\approx$3.1
AU, $e$$\approx$0.526, $i$$\approx$12$^\circ$).
The initial value of time $\tau$ when perihelion was passed was varied 
for different particles with a step $d \tau$=1 day 
(from 645 to 894$^d$) near the actual value of $\tau$ for Comet 10P at JDT 2452200.5.

For asteroidal and Encke particles, values of
the ratio between the radiation pressure force and the gravitational force
$\beta$ varied between 0.0001-0.0004 and 0.4. 
Burns et al. (1979)  obtained $\beta$=$0.573 Q_{pr}/(\rho s)$, where
$\rho$ is the particle's density in grams per cubic centimeter,
$s$ is its radius in micrometers, and $Q_{pr}$ is the radiation
pressure coefficient ($Q_{pr}$ is close to unity for particles
larger than 1 $\mu$m). For silicates,
the $\beta$ values 0.004, 0.01, 0.05, 0.1,  and 0.4 correspond to particle diameters of about 
120, 47, 9.4, 4.7, and 1 microns, respectively. 
Silicate particles with $\beta$
values of 0.01 and 0.05 have masses of $10^{-7}$ g and $10^{-9}$ g. 
For water ice, our $\beta$ values correspond to particle
diameters of 290, 120, 23, 11.7,  and 2.9 $\mu$m. 
As did Liou et al. (1999) and Moro-Martin and Malhotra (2002), 
we assume the ratio of solar wind drag 
to Poynting--Robertson drag to be 0.35. 
The relative error per integration step was taken to be less than $10^{-8}$
for asteroidal and trans-Neptunian particles and
to be less than $10^{-9}$ or $10^{-8}$ for cometary particles.
The simulations continued until all of the particles
either collided with the Sun or reached 2000 AU from the Sun. 

\section{Probabilities of collisions of dust particles with the terrestrial planets}

Orbital elements were stored with a step of 
$d_t$ of $\le$20 yr  for asteroidal and cometary particles
($d_t$=10 yr for asteroidal particles at $\beta$ equal to 0.1 
and 0.25, and $d_t$=20 yr for other runs) and of
100 yr for trans-Neptunian particles.
In our runs, planets were considered as material points, but using orbital elements 
obtained with a step $d_t$, similar to (Ipatov and Mather, 2004a) 
we calculated the mean probability 
$P$=$P_{\Sigma}/N$ ($P_\Sigma$ is the probability for all $N$ 
considered particles) of a collision of a particle with a planet 
during the lifetime of the particle. We define
$T$=$T_{\Sigma}/N$ as the mean time during which the perihelion 
distance $q$ of a particle was less than the semi-major axis of the 
planet and $T_J$ as the mean time spent in Jupiter-crossing orbits.
Below, $P_{Sun}$ is the ratio of the number of particles that
collided with the Sun to the total number of particles.
$T_S^{min}$ and $T_S^{max}$ are the minimum and maximum values of the time until
collision of a particle with the Sun, and $T_{2000}^{min}$
and $T_{2000}^{max}$ are the minimum and maximum values of  time 
when the distance between a particle and the Sun reached 2,000 AU.
The values of $P_{Sun}$, $P_r$=$10^6 P$, $T$, $T_J$, $T_S^{min}$, $T_S^{max}$, $T_{2000}^{min}$,
and $T_{2000}^{max}$  (times are in Kyr) are shown in Tables 1-4 for 
several runs with
asteroidal, Encke, Tempel 2, and trans-Neptunian particles
at different $\beta$. 

All {\it asteroidal particles} collided with the Sun at 0.004$\le$$\beta$$\le$0.01, and
$P_S$$\ge$0.96 at $0.0004$$\le$$\beta$$\le$0.1. $P_S$ is smaller for 
greater $\beta$ at $\beta$$\ge$0.1.  
The minimum time $T_S^{min}$ needed to reach the Sun is smaller for smaller particles
(i.e., for larger $\beta$). 
The ratio  $T_S^{max}/T_S^{min}$ is much greater for $\beta$$\ge$0.2 than for $\beta$$\le$0.1. 
For $\beta$=0.05, 498 of 500 asteroidal particles collided with the Sun in less than 0.089 Myr, 
but two particles (with initial orbits close to those of the asteroids
361 and 499), which reached 2000 AU, lived for 0.21 Myr and 19.06 Myr, 
respectively. The latter object's perihelion was near Saturn's orbit for a long time. 
At $\beta$$=$0.05  the first 250 asteroidal particles did not migrate outside Jupiter's orbit,
so $T_J$=0 in Table 1. 
In most runs all {\it Encke particles} collided with the Sun, but in a few runs
(e.g., at  $N$=101 and $\Delta t_\circ$=0 for $\beta$ equal to 0.4, 0.2, and 0.1)
all particles were ejected into hyperbolic orbits.


\begin{table}
\caption{Values of $T$, $T_J$, $T_S^{min}$, $T_S^{max}$, $T_{2000}^{min}$ (in Kyr),
$P_r$, and $P_{Sun}$ 
obtained for asteroidal dust particles at several values of $\beta$
(Venus=V, Earth=E, Mars=M) 
}

$ \begin{array}{llcccc ccccc ccc}

\hline

  & & &           $V$ & $V$ & $E$ & $E$ & $M$ & $M$ &&&&&\\
\beta &N&P_{Sun}&  P_r & T  & P_r & T    &  P_r & T  &T_J&T_S^{min}&T_S^{max}&T_{2000}^{min}&T_{2000}^{max}\\
\hline
0.0004&100& 0.960  &338&13.2& 375&61.4&338&225&3.68&1485&7950&336&2747\\ 
0.001 &100& 0.980  &105&9.1& 737&27.8&939  &211&1.20&407&3111&626&1230\\
0.001*&150& 0.993 &228&5.2  &342 &20.9&287&215&0.8&509&4472&884&884\\  
0.002&100& 1.000  & 2002 &48.0& 1934& 104 & 537 & 298 & 0 &592&1756&-&-\\ 
0.002*&150&0.987  & 9679 &35.3&10641&80.1 & 508 & 274 & 2 &666&2064&449&928\\
0.004&100& 1.000  &12783 &40.5&11350& 90. &1204 & 220 &0  &348&932&-&- \\ 
0.004*&150&1.000  & 2704 &38.4& 2267&81.9 & 342 & 208 &0  &338&1215&-&-\\
0.005&100& 1.000  &12207 &33.2&16700&76.6 &1020 & 184 &0  &248&1013&-&-\\
0.01 &250& 1.000  & 1534 &19.2& 1746 & 44.2 & 127 & 100&0  &142&422&-&- \\ 
0.01*&250& 0.996  & 1168 &15.2& 1269 & 33.9 & 134 & 84.8&0.02&50&422&211&211 \\ 
0.02 &250& 0.996  & 403& 9.7& 387&20.5& 52.6&48.2&0.2&65.4&178&650&650\\
0.02*&250& 0.992  & 490& 9.2& 728&19.3& 73.9&45.7&0.8&71.8&400&112&303\\
0.05 &250& 1.000  &  195 & 4.0& 190  &  8.1 &36.7 &  20&0  & 30& 89&-&- \\ 
0.1 &250&  0.988  &  141 & 2.4& 132  &  4.8 &16.4 &  12&2.21&16& 44&138&793\\ 
0.1* &250& 0.992  &  366 & 2.4& 279  &  4.8 &20.9 &  12&0.92&7.2&43&  9&534\\ 
0.2  &250& 0.852  &285 &1.6& 242  & 3.4 &21.7 &3.7&15.6&8.0&734&9.0&815\\
0.2* &250& 0.796  &642 &1.5&522& 3.1&30.8&7.3&24.7&7.2&416& 2.0&1375\\
0.25 &250& 0.618  & 79.2 & 1.4& 63.8& 2.9 & 5.60 &5.9&31.7&5.9&385&1.6&567\\ 
0.4  &250& 0.316  &  12.4& 1.5&  8.0& 2.5 & 0.72 &8.8&32.3&4.3&172&1.7&288\\

\hline
\end{array} $
\end{table}

The probability of collisions of {\it asteroidal dust particles} with the Earth was
maximum ($\sim$0.001-0.02) at 0.002$\le$$\beta$$\le$0.01, i.e., at diameters of
particles $d$$\sim$100 $\mu$m. These probabilities of collisions are in 
accordance with cratering records in 
lunar material and on the panels of the Long Duration Exposure Facility, 
which showed that the mass distribution of dust particles encountering the Earth peaks at 
$d$=200 $\mu$m (Kortenkamp and Dermott, 1998). 
For asteroidal particles with $\beta$$>$0.01, collision probabilities with the
terrestrial planets were smaller for larger $\beta$.
Values of $P$ for Venus didn't differ much from those for Earth.
At $\beta$$\ge$0.01 the values for Mars were smaller by an order of magnitude than 
those for Earth, but at $\beta$$\sim$0.0004-0.001 they were about the same.

\begin{table}
\caption{Values of $T$, $T_J$, $T_S^{min}$, $T_S^{max}$, $T_{2000}^{min}$, $T_{2000}^{max}$ (in Kyr), $P_{Sun}$, and $P_r$  for particles started from Comet Encke
(Venus=V, Earth=E, Mars=M)
}

$ \begin{array}{lcccc ccccc ccccc}

\hline

      &              &  &       &$V$&$V$&$E$&$E$&$M$&$M$&   &         &         & &\\
\beta &\Delta t_\circ&N &P_{Sun}&P_r& T &P_r&T  &P_r&T  &T_J&T_S^{min}&T_S^{max}&T_{2000}^{min}&T_{2000}^{max}\\
\hline
0.0001&0&     101&1.00&251&149&123&156&7.0&156&0&85.1&5854&-&-\\
0.0004&0&     101&1.00&319&110&125&110&9.9&110&0&83.4&679&-&-\\
0.001 & 0    &101&1.00 &257&94.0&111&94.6&9.6&94.6&0&78.8&648&-&-\\
0.002 & 0    &101& 1.00&470 &91.1& 200 & 94.6 & 12.0 &95.5& 0 &72.3&551&-&-\\
0.002 & 0    &150& 1.00&632 &92.9& 208 & 93.6 & 13.9 & 93.5&0 &75.2&370&-&-\\
0.002 & 0.25 &101& 1.00&408 &84.3& 156 & 84.3 & 12.9&84.3  &0& 80.0&208&-&-\\
0.002 & 0.5  &101&1.00& 432&86.3 & 189 & 86.3 & 13.2&86.3 & 0&80.9&240&-& -\\
0.004 & 0    &101& 1.00&370 &62.3& 148 & 62.9 & 8.9 &63.& 0 &43.7&231&-&-\\
0.004 & 0    &150& 1.00&303 &65.8& 139 & 66.0 & 9.0 &66.0&0&43.7&164&-&-\\
0.004 & 0.25 &101&1.00& 430&55.0 & 160 & 55.0 & 9.3&55.0&0&47.3 & 109&-&- \\
0.004 & 0.5  &101& 1.00&235&56.4 & 140 & 56.3 & 8.1& 56.4& 0& 45.9& 108&-&- \\
0.01  &  0  &101 & 1.00&191 & 24.9&105 & 25.1 & 5.4& 25.1&0& 17.1&67.4&-&- \\
0.01  &  0 &150 & 1.00&386&28.1 & 163 & 28.5 & 6.4 &28.5& 0&18.1&79.7&-&- \\ 
0.01  & 0.25&101 &1.00& 238&24.2 & 86  & 24.2 & 4.2&24.2 &0& 20.0&59.2&-&- \\
0.01  &0.5&251& 1.00&495&9.1&226&9.1&15.2&9.1 &0 &19.1&48.9&-&-\\
0.02  & 0 &101&0.93&413&20.8&98.3&24.1&3.7&27.1&22.4&4.5&1019&2.9&473\\
0.02  & 0 &150&1.00&89.6&13.5&37.5&13.6&1.9&13.6&0.5&8.0&426&-&-\\
0.05  & 0  &101&0.96& 12.0&7.7& 5.9&9.3&   0.6 & 11.6&23.0&2.1&527&9.4&1070\\
0.05  & 0  &150&0.99&142&7.1 & 67  & 7.8  & 2.9&8.2 &0.8&1.8&85.6&3.0&3.0\\
0.05  & 0.25&101&1.00& 37.1&4.6 & 20.5 & 4.6 & 1.6 &4.6& 0&4.4&4.7&&\\
0.05&0.5&101&1.00&96.2&6.3 & 37.2&6.4&2.3&6.4&0&5.0&20.6&-&-\\
0.1   & 0  &150&0.91& 23.1&5.2 &9.1&6.1&0.66 &7.3&3.6&1.1&112&1.1& 229 \\
0.1   & 0.25 &101&1.00& 22.4&2.8 & 8.6  & 2.8  & 0.6 &2.8& 0&2.4&3.3&-&-\\
0.1   & 0.5  &101&1.00& 13.0&2.7&   6.6& 2.7  & 0.47&2.7 &0&2.5&2.7& & \\
0.2   & 0    &150&0.60& 7.4&3.3& 3.5 & 3.6 &  0.27&3.8&3.0&1.3 & 119&0.4&3.6 \\
0.2   & 0.25 &101&1.00&20.3&1.9& 4.5 & 1.9 & 0.39&1.9 & 0&1.8&2.2&-&- \\
0.2   & 0.5  &101& 1.00&12.4&1.6 & 3.2 & 1.6 & 0.22 &1.6&0& 1.6&1.65&-&- \\
0.4   & 0.25 &101&0.58&23.6&1.3& 4.3  & 1.3  & 0.32&1.3& 0 & 1.1&1.7&1.2&1.7 \\
0.4   & 0.5  &101& 0.57&13 &1.3& 3.5  & 1.3  & 0.22 &1.3&0&1.3&1.6&1.3& 1.5\\

\end{array} $ 
\end{table}


\begin{table}
\caption{Values of $T$, $T_J$, $T_S^{min}$, $T_S^{max}$, $T_{2000}^{min}$, $T_{2000}^{max}$ (in Kyr), $P_{Sun}$, and $P_r$  for particles started from Comet 10P Tempel 2
(Venus=V, Earth=E, Mars=M)
}

$ \begin{array}{lccc ccccc ccccc}

\hline

      &                &       &$V$&$V$&$E$&$E$&$M$&$M$&   &         &         & &\\
\beta &N &P_{Sun}&P_r& T &P_r&T  &P_r&T  &T_J&T_S^{min}&T_S^{max}&T_{2000}^{min}&T_{2000}^{max}\\
\hline

0.05  & 250& 1.00& 778&5.4& 734&14.6 & 28.7&29.7&  0& 20.3&93.8&-& -\\
0.1   & 250& 1.00& 277&2.9& 246&7.2  & 18.7&17.5&  0& 11.3&30.5 &-& -\\
0.2   & 250& 0.76& 153&1.6& 114&3.7  & 7.9&7.4&  21.7& 7.4&358 &6.3& 691\\
0.4   & 250& 0.50&60.3&2.2& 44.1  & 4.0  & 3.7 &6.5&5.8&4.8&64.0&4.9& 171\\

\end{array} $ 
\end{table}


\begin{table}
\caption{Values of $T$, $T_J$, $T_S^{min}$, $T_S^{max}$, $T_{2000}^{min}$, $T_{2000}^{max}$ (in Kyr), $P_{Sun}$, and $P_r$  for kuiperoidal dust particles at $N$=50
(Venus=V, Earth=E, Mars=M). The run at $\beta$=0.01 have not yet finished.
}

$ \begin{array}{lcccc ccccc ccc}

\hline

  & & $V$ & $V$ & $E$ & $E$ & $M$ & $M$&&&&& \\


\beta & P_{Sun}& P_r & T & P_r & T &  P_r & T &T_J&T_S^{min}&T_S^{max}&
T_{2000}^{min}&T_{2000}^{max} \\ 

\hline 

0.01 &0.06& >3 & >0.4 & >3 & >0.5 & >0.6 & >0.8&>60&42,303&102,530&2,379&>177,370\\ 
0.05 &0.18& 156 & 0.18 & 134 & 0.40 & 12.6 & 1.2&16.0&5,568&18,221&2,895&50,198\\ 
0.1 &0.2&76.2 & 0.75 & 35.2  & 1.42 & 2.74 &2.8&47.4&3,659&17,439&3,730&53,949\\ 
0.2& 0.12& 182  & 0.22 & 150 & 0.46 & 13.3 &1.2&59.6&5,237&10,789&2,490&26,382\\ 
0.4 & 0.08& 44.4& 0.24 & 13.2& 0.45 &0.63&0.8&121.6&4,503&13,246&5&14,383\\

\hline
\end{array} $
\end{table}

For {\it Encke particles} colliding with Venus, Earth, and Mars, the values of 
$P$ at 0.0004$\le$$\beta$$\le$0.02 were about 0.0002-0.0006, 0.0001-0.0002,  
and (4-14)$\cdot$$10^{-6}$, respectively. They were much smaller at
$\beta$$\ge$0.05. For these planets at all $\beta$, the values of $P$ for Encke particles 
were smaller than those for asteroidal particles.
Collision probabilities of Encke particles with Earth were greater by a 
factor of 10-20 than those with Mars and greater for particles
starting at perihelion than aphelion (exclusive for $\beta$=0.4).
For the same value of $\beta$, the probability of  Encke dust particle
colliding with a terrestrial planet was less  
than for an asteroidal dust particle by a factor of several, mainly due to the 
greater eccentricities and inclinations of Encke particles (see Table 5).

The values of $T$ for {\it asteroidal dust particles} and the 
Earth were maximum ($\sim$80-100 Kyr)
at $\beta$$\sim$0.002-0.004, and they were smaller for greater $\beta$ at $\beta$$\ge$0.004.
The ratio $P_r/T$ differed  for different $\beta$ by a factor of 50.
It may be caused by that in some runs perihelia or aphelia of some particles
could be close to the orbit of the Earth or some particles moved almost in
the same plane as the Earth.
At $\beta$$\ge$0.002 for asteroidal dust particles,
the values of $T$ for Venus were about twice less than those for Earth,
and  the values for Mars were  greater 
than those for Earth by a factor of 2-3.5. 
For {\it Encke particles} the values of $T$ were
almost the same for Venus, Earth, and Mars, but they differed on 
$\Delta t_\circ$ and were greater for smaller $\beta$.  

For  $\beta$$\ge$$0.05$, the fraction $P_{Sun}$ of 
{\it trans-Neptunian particles} collided with the Sun was less
than that of {\it asteroidal particles} by a factor of 4-6.
At these values of $\beta$, collision probabilities $P$ with Earth and Venus differed
for asteroidal and trans-Neptunian particles usually by less than a factor of 2,
but the difference in $T$ was greater
(by  a factor of 3-7
at  $\beta$$\ge$$0.1$ and by a factor of 20 at $\beta$$=$$0.05$).
The mean values $e_m$ and $i_m$ of eccentricities and inclinations 
at distance $R$=1 AU from the Sun 
were mainly greater for trans-Neptunian particles than those for asteroidal particles (Table 5). 
Nevertheless, the ratio $P/T$ was greater for
trans-Neptunian particles. It may be caused by that perihelia or aphelia 
of migrating trans-Neptunian particles  more often were close to the orbit of the Earth.

  \begin{center}

\begin{table}
\caption{Mean values $e_m$ and $i_m$ (in degrees) of eccentricities and inclinations at $R$=1 AU}

$\begin{array}{lllllll}
  \hline
  $dust$ &  \beta&0.01&0.05 & 0.1 & 0.2 & 0.4 \\
\hline
$trans-Neptunian$& e_m&0.7&0.15&0.2&0.22&0.40\\
$trans-Neptunian$& i_m &16&16&13&16&21\\
$asteroidal$& e_m& 0.13 &0.09&0.12 &0.22& 0.40\\
$asteroidal$& i_m& 9.2&9.4&8.5 &9.6& 19\\
$Encke$ & e_m&0.57&0.51&0.45 &0.48&0.85\\
$Encke$ &i_m&7.5&12.8&20.9 &23.4&11.7\\
  \hline
\end{array} $
\end{table}

\end{center}

{\it Lifetimes} of particles usually were greater for smaller $\beta$.
At the same $\beta$, for Encke particles they were 
smaller by a factor of several than those for asteroidal particles.
Lifetimes of trans-Neptunian particles were $\sim$1-100 Myr (Table 4),
i.e., were much greater than those of asteroidal particles.
The run at $\beta$=0.002 have not yet finished, and 
at 3.4 Myr all trans-Neptunian particles were still moving
in elliptical orbits.
 
Liou et al. (1996) noted that interstellar dust particles with 
an average size of 1.2 $\mu$m
can destroy dust particles formed in the solar system
and that the collisional lifetimes for 1, 2, 4, 9, 23 $\mu$m particles 
are 104, 49, 19, 4.8, 0.86 Myr, respectively.
In these size ranges mutual collisions are not as important as collisions
with interstellar grains.
Moro-Martin and Malhotra (2002) concluded that {\it collisional destruction}
is most important for kuiperoidal grains between 6 $\mu$m 
(9 $\mu$m in Liou et al., 1996) and 50 $\mu$m. Particles larger than 50 $\mu$m may survive 
because interstellar grains are too
small to destroy them in a single impact. 
Taking into account lifetimes of trans-Neptunian particles presented in Table 4,
we can conclude that most of 1-8 $\mu$m silicate particles can reach the Sun without destruction,
but the fraction of larger particles destroyed during the 
motion to the Sun can be considerable.
As the mass of the trans-Neptunian belt is greater 
than the mass of the asteroid belt
by more than two orders of magnitude, and the values of $T$ in Table
1 are greater by less than a factor of 20 than those in Table 4 at the same $\beta$ for $\beta$$\ge$0.05, then for $d$$\sim$1-10 $\mu$m the fraction of 
trans-Neptunian particles among particles from different sources can be 
considerable even at $R$$<$3 AU, but they are not icy,
as icy trans-Neptunian particles evaporate before they reach the near-Earth space.
Liou et al. (1996) and Moro-Martin and Malhotra (2002) noted that 
for silicate
particles 1-40 $\mu$m in diameter, the sublimation temperature ($\sim$1500 K) is
reached at $R$$<$0.5 AU, but for water ice particles the sublimation temperature 
($\sim$100 K) is reached at 27, 19, 14, 10, and 4.3 AU for the sizes of 3, 6, 11, 23, 
and 120 $\mu$m, respectively.

For particles started from Comet 10P, the obtained results were closer to those for particles started from asteroids than from Comet 2P Encke. In considered runs
the values of $T$ for Comet 10P were even greater than those for asteroids. 

For asteroidal particles and the {\it model without planets}, the values of $T_S^{min}$
were about the same as those presented in Table 1,
but the values of $T_S^{max}$ sometimes were smaller (Ipatov et al., 2004a). With $N$=250 
the values of $P_{Sun}$ for $\beta$=0.25 and $\beta$=0.4 of 0.908 and 0.548,
respectively, are greater than for the model with planets.
For  $\beta$$\le$0.1 all of the particles collided with the Sun.

\section{Migration of dust particles}

At $\beta$$\le$0.1 most of asteroidal and cometary particles 
didn't migrate outside Jupiter's orbit. Several plots of the distribution of 
migrating asteroidal particles in their orbital elements and  the distribution 
of particles with their distance $R$ from the Sun and their height $h$ above the initial 
plane of the Earth's orbit were presented by Ipatov et al. (2004a).
For all considered $\beta$, the mean time $t_a$ (the total time divided by the number $N$ of particles) during which an asteroidal dust particle (below in this paragraph
we consider {\it asteroidal particles}) had 
a semi-major axis $a$ in an interval of fixed width
decreases considerably  with a decrease of $a$ at $a$$<$1 AU, and
it is usually greater for smaller $\beta$ at $a$$<$3 AU. 
For $\beta$$\le$0.1 the values 
of $t_a$ are much smaller at $a$$>$3.5 AU than at $1$$<$$a$$<$3 AU, and the local maxima of $t_a$ 
corresponding to the  6:7, 5:6, 3:4, and 2:3 resonances with the Earth
are greater than the maximum at $a$$\approx$2.3 AU. 
There are several other local maxima  corresponding to the n:(n+1) resonances 
with Earth and Venus (e.g., the 7:8 and 4:5 resonances with Venus).
The trapping of dust particles in the n:(n+1) resonances cause Earth's asteroidal ring 
(Dermott et al., 1994a-b). 
Ipatov et al. (2004a) showed that
the greater the $\beta$, the smaller the local maxima corresponding to these resonances.
At $\beta$$\le$0.1 there are gaps with $a$ a little smaller than the semi-major 
axes of Venus and Earth that correspond to the 1:1 resonance for each; the greater the 
$\beta$, the smaller the corresponding values of $a$. 
A small gap for Mars is seen only at $\beta$$\le$0.01.
There are also gaps corresponding to the 3:1, 5:2, and 2:1 resonances
with Jupiter.
At $\beta$=0.01 some asteroidal particles migrated into the 1:1 resonance  with Jupiter.
For $a$$>$10 AU perihelia were usually near Jupiter's orbit, but sometimes 
also near Saturn's orbit. 


The mean time spent by an asteroidal dust particle in inner-Earth
($Q=a(1+e)$$<$0.983 AU), 
Aten 
($a$$<$1 AU, $Q$$>$0.983 AU), 
Apollo ($a$$>$1 AU, $q=a(1-e)$$<$1.017 AU), and Amor
($a$$>$1 AU, 1.017$<$$q$$<$1.300 AU) orbits was 
22, 5.6, 18, and 30 Kyr at $\beta$=0.01
and 0.4, 0.3, 1.8, and 3.0 Kyr at $\beta$=0.4, respectively.
The ratio of mean times spent by Encke particles in inner-Earth, Aten, and Apollo 
orbits was about 1.5 : 1 : 2, but varied from run to run.
The total mass of comets inside Jupiter's orbit
is much smaller than the total mass of asteroids, but a comet produces more dust 
per unit minor body mass than an asteroid.

Analysis of the Pioneer 10 and 11 meteoroid detector data 
(Humes, 1980; Gr\"un, 1994)  showed
that a population of $10^{-9}$ and $10^{-8}$ g ($\sim$10 $\mu$m)
particles has a constant spatial density between 3 and 18 AU. 
The spatial density of 1.4-10 $\mu$m particles obtained 
with Voyager 1 data was constant from 30 to 51 AU.  
Liou et al. (1999) and Liou and Zook (1999) concluded that dust grains released by 
Halley-type comets cannot account for this 
observed distribution, 
but trans-Neptunian dust particles can.

In our runs, beyond Jupiter's orbit even the number $n_R$  of asteroidal 
particles at some distance $R$ from the Sun is smaller for greater $R$,
and a {\it spatial density} $n_s$ is proportional to $n_R/R^2$.
For asteroidal and cometary particles, $n_s$ quickly decrease with an increase of $R$,
e.g., for $\beta$=0.2,  $n_s$ was smaller at $R$=5 AU than at $R$=1 AU
by a factor of 70 and 50 for asteroidal and Encke particles, respectively. 
So asteroidal dust particles cannot explain the constant spatial density of dust 
particles at $R$$\sim$3-18 AU. 
At such distances, many of the dust particles could have come from the trans-Neptunian
belt or from passing comets.
In our runs at $\beta$$\ge$0.05 a spatial density $n_s$ of considered 
trans-Neptunian particles near ecliptic
at $R$=1 AU was greater than at $R$$>$1 AU. 
At 0.1$\le$$\beta$$\le$0.4 and 2$<$$R$$<$45 AU 
(at $\beta$=0.05 for $11$$<$$R$$<$50 AU) for trans-Neptunian particles, $n_s$ varied with $R$ by less than a factor of 4.
This result is in agreement with the observations and with
the simulations made by Liou and Zook (1999) with the use of RADAU integrator.

The spatial density of  dust particles was greater for smaller $R$ at $R$$<$4 AU 
(exclusive for Encke particles at $\beta$=0.4).
For asteroidal and trans-Neptunian dust, depending on $\beta$, it was 
more at 1 AU than at 3 AU by a factor of 2.5-8 (by a factor of 10-16 for 
Encke particles at 0.01$\le$$\beta$$\le$0.2). 
This is in accordance with the observations for the inner solar system:  
inversion of zodiacal light observations by the Helios spaceprobe 
revealed a particle density $n_s$$\propto$$R^{-1.3}$,
Pioneer 10 observations between the Earth's orbit and the asteroid belt yielded 
$n_s$$\propto$$R^{-1.5}$, and IRAS observations have yielded $n_s$$\propto$$R^{-1.1}$ 
(Reach, 1992). 

\section{Brightness of dust particles}

Ipatov et al. (2004b) investigated how the solar spectrum is 
changed by scattering by dust particles (a detailed paper will be prepared).   
Positions of particles were taken from the runs discussed above.
For each such 
stored position, we calculated  $>$10$^3$ different positions of a particle and the 
Earth during the period $P_{rev}$ of 
revolution of the particle around the Sun,
considering that orbital elements do not vary during $P_{rev}$. 
Three different scattering functions
were considered. In the first model,
the scattering function depended on a scattering angle $\theta$ in such a way: 
1/$\theta$ for $\theta$$<$c, 1+($\theta$-c)$^2$ for $\theta$$>$c, 
where $\theta$ is in radians and c=2$\pi$/3 radian. In the second model, we added the same dependence 
on elongation $\epsilon$ (considered westward from the Sun). 
In the third model, the scattering function didn't depend on 
these angles at all. For all these three models, the scattering function
was proportional to $\lambda^2$$\cdot$$(R*r)^{-2}$, where $r$ is the distance between
a particle and the Earth and $\lambda$ is a wavelength of light. For each considered position, we calculated
velocities of a dust particle relative to the Sun and the Earth and
used these velocities and the scattering function for construction 
of the solar spectrum received at the Earth after been scattering
by different particles located at some beam (view of sight) from the Earth.
The direction of the beam is characterized by  $\epsilon$ and inclination $i$.
Particles in the cone of 2$^\circ$ around this direction were considered.
In each run, particles of the same size (at the same $\beta$)
and the same source (i.e., asteroidal) were studied.

The plots of the obtained spectrum (e.g., Fig. 1) are in general agreement with the 
observations made by Reynolds et al. (2004) who 
measured the profile of the scattered solar Mg I$\lambda$5184 absorption 
line in the zodiacal light.
Unlike results by Clarke et al. (1996), our modeled spectra don't exhibit strong asymmetry. 
As these authors, we obtained that minima in the plots of dependencies 
of the intensity of light on its wavelength near 5184 Angstrom are not so 
deep as those for the initial solar spectrum. The details 
of plots depend on diameters, inclinations, and a source of particles.
Different particles 
populations produce clearly distinct model spectra of the zodiacal light. 
For example, for  $i$=0  and kuiperoidal 
particles, the shift of the plot to the blue  was greater 
than those for asteroidal and Encke particles at  $\epsilon$=90$^\circ$,
and the shift to the red was greater at $\epsilon$=270$^\circ$.
The results of modeling are relatively 
insensitive to the scattering function considered, the difference was greater
for more close direction to the Sun.
Our preliminary models and comparison with observational data indicate that 
for more precise observations it will be possible to distinguish well 
the sources of the dust and impose constrains on the particle size. 

\begin{figure}
\includegraphics[width=125mm]{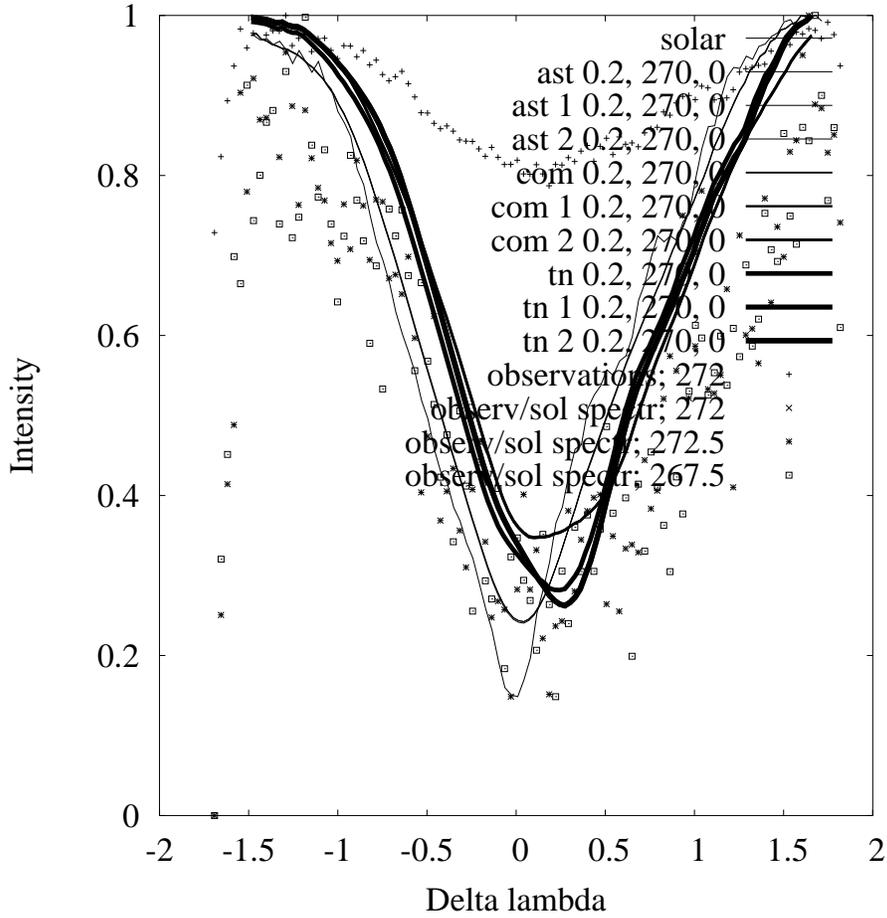}

\caption{Dependence of the intensity of light vs its wavelength $\lambda$ 
(in Angstrom) at $\beta$=0.2, $\epsilon$=270$^\circ$, and $i$=0.
Zero of $\Delta \lambda$=$\lambda$-$\lambda_\circ$ corresponds to 
$\lambda$=$\lambda_\circ$=5184 Angstrom.
The most thin line  denotes the initial solar spectrum.
Solar spectra for asteroidal ('ast') and Encke ('com') particles are
practically the same for three
scattering functions. For trans-Neptunian ('tn') particles for the first and the second
models (e.g., denoted as 'tn 1' and 'tn 2', respectively) the plots are practically the same,
but the plot for the third model (denoted simply as 'tn') is different. 
For observations (made by Reynolds et al., 2004) only 
the value of elongation is presented in the legend.
Designation "observ/sol spectr" corresponds to the case
for which the plot based on the observations was stretched in such a way that
the minimum became the same as that for the initial solar spectrum.
The maximum value was considered to be the same (equal to 1) for all plots.
} 
\end{figure}%

For asteroidal and Encke particles at $i$=0, about 65-89\% and 70-85\%  of brightness 
was due to the particles at distance from the Earth $r$$<$1 AU
(83-96\% for $r$$<$1.5 AU; 80-98\% for $R$$<$2 AU). For trans-Neptunian particles,
14-78\%, 22-85\%, 26-78\%, and 40-90\% of brightness was due to  $r$$<$1 AU,
$r$$<$1.5 AU, $R$$<$2 AU, and $R$$<$5 AU, respectively. The above ranges 
were caused by different values of $\beta$ and $\epsilon$ and different
scattering functions considered. Only a few trans-Neptunian particles in one run reached
the near-Earth space, so statistics was not good and could increase the above intervals.
According to Gr\"un (1994), the intensity $I$ of zodiacal light falls off with heliocentric distance
$R$ as $I$$\sim$$R^{-\gamma}$, with $\gamma$=2 to 2.5 and beyond about 3 AU
zodiacal light was no longer observable above the background light. 
At $\beta$$\ge$0.05 
the brightness of all trans-Neptunian dust particles located at $R$$>$3 AU
was less by only a factor of several than that at $R$$<$3 AU,
so the contribution of trans-Neptunian
dust particles at $d$$<$10 $\mu$m to the zodiacal light may not be large
(else zodiacal light will be observed beyond 3 AU), but this problem needs more accurate estimates. Based on our runs, we suppose that the fraction
of trans-Neptunian dust particles among particles of different origin for
larger particles can be much smaller than those for $d$$<$10 $\mu$m. Note that it is
considered that
the main contribution to the zodiacal light is from particles with 
diameters of about 20 to 200 $\mu$m. 

Velocities of dust particles relative to the Earth that mainly 
contributed to brightness were different for different $\epsilon$. At $i$=0 
they were between -25 and 25 km/s.

\section{Conclusions}

Some Jupiter-family comets (JFCs) can reach 
typical near-Earth object (NEO) orbits and remain there for millions of years. 
While the probability of such events is small ($\sim$0.1 \%), nevertheless the 
majority of collisions of former JFCs with the terrestrial planets are due 
to such objects. Most former TNOs that have typical NEO orbits moved in such 
orbits for millions of years, so during most of this time there were extinct 
comets. If those former JFCs that got NEO orbits for millions of years didn't 
disintegrate during this time, there could be many 
(up to tens of percent) extinct comets among  NEOs.

Collision probabilities of migrating asteroidal and
cometary dust particles
with the terrestrial planets during the lifetimes of these particles
were maximum at diameter $d$$\sim$100 $\mu$m, 
which is in accordance with the analysis of microcraters.
The probability of collisions of cometary particles with the Earth
is smaller than for asteroidal particles, and this difference is greater
at  $d$$\sim$100 $\mu$m.
At $d$$<$10 micron, 
the mean time spent by a former trans-Neptunian particle
in NEO orbits   is less 
than that for an asteroidal dust particle by
an order of magnitude, and the difference
in collision probabilities with the Earth during a lifetime of a particle is less
than the difference in the mean time.

\section*{Acknowledgements}

     This work was supported by NASA (NAG5-12265) and INTAS (00-240).


\end{document}